\documentstyle[12pt,epsf,a4wide,./mathsymb]{article}
\begin{document}

\title{
\begin{flushright}
{\normalsize SUSX-TH-96-012}\\[-3mm]
{\normalsize \tt hep-lat/9608085}\\[1cm]
\end{flushright}
Improvements for Vachaspati-Vilenkin-type Algorithms for Cosmic String
and Disclination Formation}
\author{Karl Strobl\\
{\it Centre for Theoretical Physics,}\\
{\it University of Sussex, Brighton BN1 9QH, U.K.}}
 
\date{August 1996\\[4mm]
{\small\tt PACS: 02.70.Rw 11.10.Lm 61.72.Lk }}
\maketitle

\begin{abstract}
We derive various consistency requirements for Vachaspati-Vilenkin type
Monte-Carlo simulations of cosmic string formation \cite{VV}
or disclination formation in liquid crystals \cite{MHKSII}. We argue for
the use of a tetrakaidekahedral lattice in such simulations.

We also show that these calculations can
be carried out on lattices which are formally infinite, and
do not necessitate the specification of any boundary conditions.
This way string defects can be traced up to much larger lengths than
on finite lattices.

The simulations then fall into a more general class of simulations
of self-interacting walks, which occupy the underlying lattice very
sparsely. An efficient search algorithm is essential. We discuss various
search strategies, and
demonstrate how to implement hash tables with collision
resolution by open addressing, as used in \cite{MHKSII}.
The time to trace a string defect is then proportional only to
the string length.
\end{abstract}

\newpage

Line defects are formed after a phase transition if the manifold
of equilibrium states ${\cal M}$ (the vacuum manifold) is not simply connected
\cite{Kibble-mechanism}.
They are studied theoretically in the context of field
theories in the early Universe under the name of cosmic strings,
but also exist in the laboratory in the form of superfluid
vortices, super--conductor flux tubes, and
line disclinations in nematic liquid crystals. The
interest in the formation of defects in cosmological phase transitions
has also been reflected in laboratory experiments studying the formation
and evolution of defects in nematic liquid crystals \cite{Che+90,Bowick}
and He-II \cite{McClin} (for a comprehensive review of such experiments
see ref.~\cite{Zurek}).

The properties of ensembles of
one-dimensional objects have importance in many problems in physics:
polymer science \cite{deG}; dislocation melting \cite{Disloc}; the
liquid-gas transition \cite{Riv}; and the Hagedorn transitions in
effective \cite{Hag} and fundamental \cite{Sundborg,BowWij,MitTur}
theories of strings at high temperature.
In refs.~\cite{MHKS,MHKSII} we presented Vachaspati-Vilenkin type
Monte Carlo simulations of cosmic string formation. The algorithms
used in those measurements
contained many improvements to the usual Vachaspati-Vilenkin
algorithm on finite cubic lattices \cite{VV,V91,Allega,KibZ2}.

In this paper we present the details of these improvements.
We suggest the use of a tetrakaidekahedral lattice, to ensure uniqueness
in tracing the shape of a string defect.  We show that Vachaspati-Vilenkin
calculations can be reduced to a treatment of the string defects
as self-interacting walks, independent of the neighbouring walks.
In this case, one needs to store only the walk coordinates
in the computer memory, and a very sparsely populated
lattice results. This suggests the use of hash tables to search for
occupied lattice sites. We explain how hash tables with collision
resolution by open addressing have been implemented for the
work of ref.~\cite{MHKSII}.

Similar search algorithms -- discussed in section \ref{section:sortlist} --
have been used in simulations of string
dynamics \cite{EPSS,BB}, where the lattice has also been sparsely
populated by strings.
In the simulations for string {\it formation} one can, however, take
advantage of two additional facts: firstly, we do not take account of the
string dynamics, such that no data base entry has to be deleted, allowing
us to use the slightly more efficient collision resolution by open
addressing. Secondly, the string defects can be interpreted as mutually
non-interacting, such that the lattice can be made
formally infinite and no (unphysical) boundary condition has to be
imposed in order to insure numerical tractability of the problem.
This realisation is inspired by Monte Carlo techniques to measure the
statistics of self-avoiding walks (see e.g.~ref.~\cite{Sokal} and
references therein), but is valid for any mutually non-interacting walks.

In section \ref{section:1} we define the Vachaspati-Vilenkin
construction of string defects for general lattices and ground state
symmetries of the underlying field theory.
In section \ref{section:2} we show that this algorithm has only very
unrestrictive requirements in order to preserve both
the topological definition of a string defect and satisfies string
flux conservation in the lattice prescription.
In section \ref{section:3} various advantages of using the dual to the
tetrakaidekahedral lattice are presented.
Finally, section \ref{section:4} introduces a numeric representation of
the Vachaspati-Vilenkin strings which allows to consider a formally
infinite lattice. An efficient data structure and search algorithm
are needed to put this representation to practical use. We introduce
various different possible data structures, and show that there is
one which allows to trace a string in a computational time of the order
of the string length only.
In this method, we build up a lattice
as it is needed to trace the walk, and entirely disregard lattice sites
which are never visited by the string defect or dislocation. We can build
up the lattice ad infinitum, and computer memory is only needed to store
the coordinates of a {\it single} walk, allowing us to trace much longer
strings than possible on finite size lattices, without ever running into
a lattice boundary.
The realisation of this method on a bcc lattice, used in ref.~\cite{MHKSII}
is presented as an example.

\section{The Vachaspati-Vilenkin algorithm}
\label{section:1}

A topological line-defect, i.e.~a cosmic string, a line disclination
in a nematic liquid crystal (NLC), a vortex line in superconductors,
or a vortex in liquid He, has a shape determined by the field
map from 3-space into the ground state manifold after a phase transition,
if the ground state manifold ${\cal M}$ (or vacuum manifold) has a non-trivial
first homotopy group $\pi_1({\cal M})$, i.e.~if it contains non-contractable
loops.
The appropriate ${\cal M}$-symmetric
field undergoing the phase transition is uncorrelated
beyond a certain scale $\xi$. If
a closed path in space is mapped -- through the field map --
onto a non-contractable loop on ${\cal M}$, the spatial path encloses
a string defect, which is topologically stable\footnote{For an excellent
review article, the reader may want to consult ref.~\cite{MarkTom}.}.
In cosmology this is known as the Kibble mechanism \cite{Kibble-mechanism}.
Tracing a string initially is a matter of knowing
the particular field value at each space-point close enough to the string.
For the late-time dynamical simulations, it is permissible and more practical
to consider the string as an infinitely thin
classical object with its mass and string
tension as the only relevant parameters \cite{EPSS,BB,Graham,AT},
although the Vachaspati-Vilenkin method is commonly used to give the
initial conditions of the string network to evolve in such simulations.

To model the Kibble mechanism, the lattice spacing in the Vachaspati
Vilenkin (VV) algorithm is interpreted as $a\stackrel{>}{\sim}\xi$, so that
random field values $\phi\in{\cal M}$ can be
assigned to each lattice point.
In a second order transition, $\xi$ is just the
Compton wavelength of the scalar particle, while in a first order
transition, such as in a rapid quench in a liquid crystal, it is interpreted
as the average bubble separation~\footnote{At first sight, it appears
more appropriate to use a random lattice for the first order case. This
is not necessary, however, because of the observed universality \cite{MHKSII},
as long as general symmetries of the problem
(e.g.~rotational symmetry and flux conservation)
are conserved by the lattice. The large distance behaviour will then be
independent of the lattice structure. We concern ourselves with the
conservation of these physical symmetries in the next section.}.
To see where the strings will be located initially, one connects the
field values along the spatial links by using the geodesics on the
vacuum--manifold which connect the field values assigned to the end points
of the link~\cite{RudSri,Hin+}.
It is reasonable to assume that the field will fall into this
configuration because this minimises the gradient--energy in the field. Along
any closed path consisting of consecutive lattice links, it is then easy to
demonstrate that the total string flux in any area enclosed by this path is
given by the number that specifies the homotopy class containing the resulting
closed walk image in ${\cal M}$.
In this paper, we do not restrict the definition of the VV algorithm to any
particular lattice, to any particular manifold ${\cal M}$, or to particular
discretisations thereof.

\begin{figure}[htb]
\centering
\mbox{~}{\hbox{
\epsfxsize=300pt
\epsffile{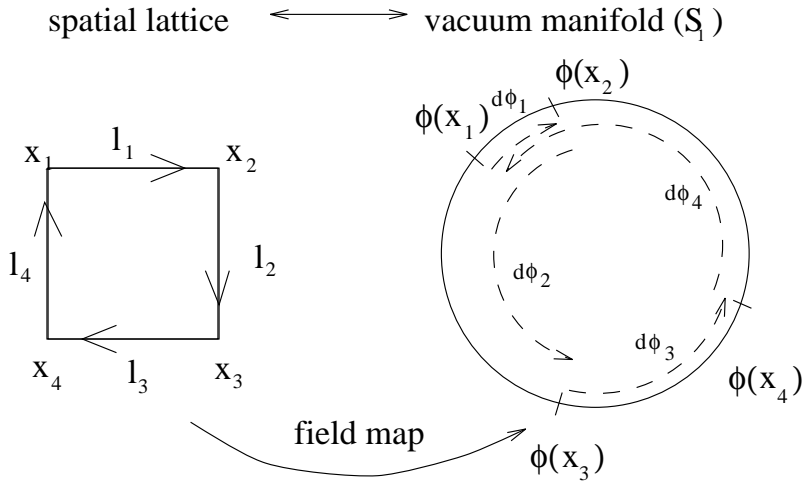}}}\\
\caption{Schematic illustration of the VV algorithm for
a $U(1)$ manifold on a cubic lattice. The
field values $\phi\in[0,2\pi)$
are assigned randomly, since the link lengths correspond
to one correlation length in the field. The field along the links follows
the shortest possible paths (geodesics) to connect between the field values
at the end points of the links. In the case at hand, the field winds around
counterclockwise as we follow the links $l_1$ to $l_4$, so the winding
number is -1, and the quadratic face will be penetrated by a string coming out
through the face (or going in, depending on the convention for the string
orientation).}
\label{fig:VV}
\end{figure}

In the simple case of a $U(1)$ symmetry on a cubic lattice for example,
the total string flux through a lattice plaquette is just the number
of times the series of geodesics on the vacuum manifold
(corresponding to the field map of the lattice links)
winds around the $U(1)$ manifold. Taking an elementary
quadratic face of the cubic lattice, one can get winding numbers of
-1, 0, or 1, so that one can just identify the nonzero numbers with the
existence of a string somewhere through this face.
This concept is sketched in Fig.~\ref{fig:VV}.

\section{VV Algorithms and Flux Conservation}
\label{section:2}

Under fairly unrestrictive conditions -- which we investigate below --
this algorithm has two important properties:
\begin{enumerate}
\item The string flux through adjacent lattice areas
obeys the addition defined through $\pi_1({\cal M})$.
This means that the string flux through any
lattice area is equal to the topological
winding number of the mapping of a walk along the bounding lattice links,
but {\it also} equal to the $\pi_1({\cal M})$-sum of the string
fluxes through the constituting lattice plaquettes.
For $U(1)$ strings for example,
the (quantised) flux $n$ through a lattice surface
$S$ can be expressed by the changes of the $U(1)$ angular parameter
$\theta$ when walking along all links of the boundary $\partial S$
of $S$:
\begin{equation}
n_S=\frac{1}{2\pi}\sum_{l\in \partial S}d\,\theta_l
\,\,\,\,\,.\,\,\,\,\,
\label{eq:Stokes}
\end{equation}
where $l$ has an orientation as well as a position on the lattice.
From this equation we also see that the flux for {\it any} two adjacent
surfaces must be additive, as long as
\begin{equation}
d\vec{\theta}_l=-d\vec{\theta}_{-l}\,\,\,\,\,,\,\,\,\,\,
\label{equation:Abel}
\end{equation}
which is generally a very unrestrictive requirement, which can be satisfied
in several ways, and which reflects the vectorial structure
of the field gradients in the continuum picture.
This means that even the geodesic rule may be broken \cite{PaulSaffin},
without affecting
the topological definition of strings in the VV algorithm.
The proof is now straightforward: if we form one surface $S$ from
the two adjacent ones $S_1$ and $S_2$,
the sum of walks in both directions along
the parts of the edges that disappear cancels out.
In the following we will assume that $\{S_1,S_2\}$ is a partition of
$S$, i.e.~$S_1\cap S_2=\emptyset$ and $S_1\cup S_2=S$. We should have
\begin{equation}
n_S=n_{S_1} + n_{S_2}=
\frac{1}{2\pi}\left[\sum_{\partial S_1}d\,\theta+\sum_{\partial
S_2}d\,\theta\right]
\,\,\,\,\,,\,\,\,\,\,
\label{equation:3}
\end{equation}
The parts that $\partial S_1$ and $\partial S_2$ have in common are
traversed in both directions, when calculating the sum of string fluxes,
so that, according to equation \ref{equation:Abel}, the ``inner''
contributions to
the total string flux cancel out, and only the outer parts of the
boundary contribute. Thus, we have
\begin{equation}
\sum_{\partial S_1}d\,\theta+\sum_{\partial S_2}d\,\theta=
\sum_{\partial S}d\,\theta\,\,\,\,\,.\,\,\,\,\,
\label{equation:flux}
\end{equation}
This means that {\it
Eq.~\ref{equation:Abel} ensures that the string flux survives in its
lattice definition as a topological quantity.}
The proof is trivially generalised to other vacuum manifolds ${\cal M}$,
by simply letting $\theta$ be in the representation of a parameterisation
of ${\cal M}$, but the addition $\oplus$ defined through the group
$\pi_1({\cal M})$ has to be used in Eqs.~\ref{equation:3} and
\ref{equation:flux}, since any
$\sum_{\partial S}d\,\theta$ is an element of this group, if
$\partial S$ is the boundary of a surface $S$, i.e.~if $\partial S$ is
a (collection of) closed walk(s).
We adopt this general notation from now on.
\item {\it Flux conservation} is another direct consequence of
Eq.~\ref{equation:Abel}. The general form of Eq.~\ref{equation:3}
is
\begin{equation}
n_S=n_{S_1}\oplus n_{S_2}\,\,\,\,\,,\,\,\,\,\,\mbox{if~}S=S_1\cup S_2, S_1\cap S_2=\emptyset
\,\,\,\,\,,\,\,\,\,\,
\label{equation:addition}
\end{equation}
where $\oplus$ is the addition appropriate for the group $\pi_1({\cal M})$,
i.e.~for example the usual addition of integers in the case ${\cal M}=U(1)$,
or the addition of $\Z_2$ for so-called $\Z_2$ strings like e.g.~strings
with ${\cal M}=\R P^2$.
One consequence of
Eq.~\ref{equation:addition} is that no strings terminate except by
running into themselves and forming closed loops:
for any {\it closed} surface $S$ and any partition thereof into
two surfaces $S_1$ and $S_2$,
$\partial S_1$ and $\partial S_2$ are identical except for their
orientation, and the total string flux through $S$ is therefore
\[
\sum_{\partial S_1}d\,\theta\oplus
\sum_{\partial S_2}d\,\theta=
\sum_{\partial S_1}d\,\theta\oplus
\sum_{\partial S_1}(-d\,\theta)=0\,\,\,,
\]
as long as Eq.~\ref{equation:Abel} is satisfied.
Thus we conclude that {\it Eq.~\ref{equation:Abel} also ensures
that the strings in the VV algorithm do not terminate anywhere unless they
close back onto themselves to form closed loops.}
\end{enumerate}

\section{The Lattice Structure}
\label{section:3}

\begin{figure}[htb]
\centering
\mbox{~}{\hbox{
\epsfxsize=120pt
\epsffile{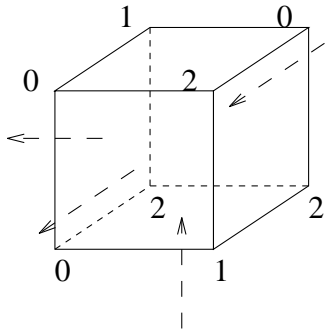}}}\\
\caption{A case where a cube is penetrated by two strings. Let the numbers
correspond to, say, $\phi= 0,\,\,2\pi/3,\,\,\mbox{and }4\pi/3$. Then the
bottom and the back face are letting pieces of string in, while the front
and the left face have pieces of string leaving the cube. There
is no obvious way to uniquely connect a particular incoming piece with an
outgoing one.}
\label{fig:cubic_problem}
\end{figure}

We have now shown that  -- after assigning random field values to
each lattice point -- we can uniquely
define which elementary faces are penetrated by a string without running
into problems with the topological nature of strings or the infinite
volume limit. This is valid for {\it any} lattice and {\it any} vacuum
manifold, as long as Eq.~\ref{equation:Abel} is satisfied, but only
concerns the string flux.
The problem with combining particular lattices with certain
vacuum manifolds, like e.g.~a cubic lattice where the
manifold is $U(1)$, is one of uniqueness, not in terms of
a position of some short piece of string penetrating an elementary face,
but in the way those short pieces connect to form distinctly
infinite strings or string loops.
A case of this happening is depicted in Fig.~\ref{fig:cubic_problem}, where
a cube is penetrated by two strings, and it is not clear how the ends are to
be connected within the cube.
Random reconnection of the free ends has often been used as a way to
resolve this problem, but this introduces an unphysical bias towards Brownian
statistics on large scales (i.e.~on scales where, on average, a large number
of such ambiguities will arise).

\subsection{Tetrahedral Lattices}

For many vacuum manifolds (or discretisations thereof),
a tetrahedral lattice avoids such ambiguities.
Effectively, we have to obtain a proof, for any new representation of
${\cal M}$, that the string defect will be a self-avoiding walk on the
dual lattice\footnote{Note that we do not mean a self-avoiding {\it random}
walk, which is the subject of entirely different research. The self-avoidance
of the string defects in the VV algorithm is not of a random nature, but
dictated by the topology of the field. It is the {\it field} values on
the lattice sites which are random (and in fact, quite unlike the walk
segments of the self-avoiding random walk, Markovian).},i.e.~no tetrahedron
can ever bear more than one string.

Let us give two examples for such proofs:
If we discretise the vacuum manifold ${\cal M}=U(1)$ minimally,
i.e.~by three
points with equal spacing $d\,\theta=2\pi/3$ between them,
a tetrahedron lets at most one string pass through it.
\begin{list}{{\it proof:}}{}
\item
For a string to pass through a particular triangular face, the vertices
of the face all have to have different vacuum angles. No
triangular face where any two of the three vertices have the same phase
can carry a string. If ${\cal M}$ is discretised by three points, at least two
points of the tetrahedron must have the same field value, i.e.~at least
two triangles cannot carry a string.\mbox{~~}{\Large $\Box$}
\end{list}
As it turns out, there is a proof for the full continuous $U(1)$
symmetry that a tetrahedron cannot have elements of $U(1)$ assigned
to its vertices in such a fashion to make it carry more than one string
inside (if the geodesic rule is to hold). This proof trivially
extends to any discretisation of $U(1)$.
\begin{list}{{\it proof for continuous $U(1)$:}}{}
\item Say that two vertices are
assigned the phases $\phi_1$ and $\phi_2$. In order for the triangle
$(x_1,\,x_2,\,x_3)$ to carry a string, 
$\phi_3$ has to be in the interval $\Gamma_{12}$,
indicated in Fig.~\ref{fig:Gamma},
i.e.~$\phi_3$ has to be between the
angles $\overline{\phi_1}=(\phi_1+\pi)\,\,{\rm mod}\,\, 2\pi$ and
$\overline{\phi_2}=(\phi_2+\pi)\,\,{\rm mod}\,\, 2\pi$
(``between" means in the shorter
of the two ways this interval can be defined on a circle, which is the
interval that does not include $\phi_1$ and $\phi_2$).
\begin{figure}[ht]
\centering
\mbox{~}{\hbox{
\epsfxsize=120pt
\epsffile{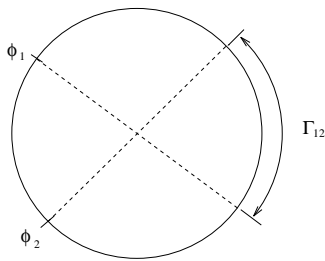}}}\\
\caption{The set $\Gamma_{12}$ of values $\phi_3$ which would -- according to
the geodesic rule -- form a string if such a $\phi_3$ occurred together with
$\phi_1$ and $\phi_2$ on the vertices of a triangle.}
\label{fig:Gamma}
\end{figure}
By the symmetry of the problem, it follows not only that $\phi_1\in\Gamma_{23}$
and $\phi_2\in\Gamma_{13}$, but also that
$\Gamma_{12}\cup\Gamma_{13}\cup \Gamma_{23}=S^1={\cal M}$,
and that all sets $\Gamma_{ij}$
have no overlap of non-zero
measure, i.e.~$\|\Gamma_n\cap\Gamma_m\|=0\,\,,\,\,{\rm if~}m\ne n$
(cf.~Fig.~\ref{fig:proof2}).
\begin{figure}[ht]
\centering
\mbox{~}{\hbox{
\epsfxsize=120pt
\epsffile{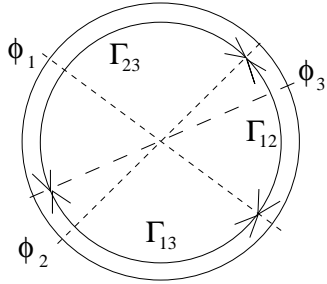}}}\\
\caption{The sets $\Gamma_{ij}$ form a complete partition of the circle.
All possible values for $\phi_4$ can therefore
form a string only with exactly two other values of $\phi$, i.e.~on exactly
one additional face of the tetrahedron.}
\label{fig:proof2}
\end{figure}
Any fourth vertex
of the tetrahedron therefore has to have a field value within exactly one
$\Gamma_{ij}$~\footnote{Unless
$\phi_4$ is exactly on one of the borderlines. Since
any numerical representation of $U(1)$ is a (although very fine) discretisation,
it is easy to ensure that such points do in fact not occur, by choosing for
instance a discretisation of $U(1)$ in terms of a very large but odd number
of equally spaced points.}, which means that it will form a string on exactly
one other tetrahedron--face, namely the face
$(n,m,4)\,\,,\,\,{\rm if}\,\,\, \phi_4\in\Gamma_{nm}$.
\mbox{~~}{\Large $\Box$}
\end{list}
For a minimal discretisation of an  $\R P^2$ manifold the proof is shown
in ref.~\cite{MHKSII}. However, one also finds that this uniqueness of
the string shapes gets lost -- even on a tetrahedral lattice -- if the
vacuum manifold has an $\R P^2$ symmetry with a continuous representation.

\subsection{The Tetrakaidekahedral Lattice}

\begin{figure}[ht]
\centering
\mbox{~}{\hbox{
\epsfxsize=330pt
\epsffile{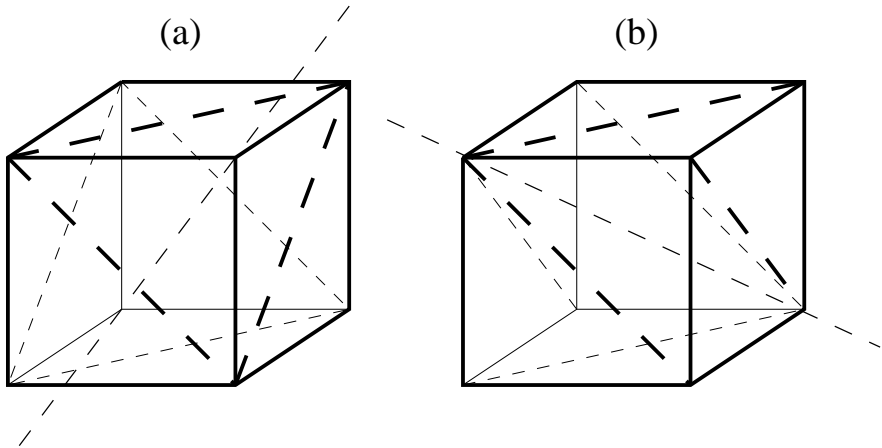}}}\\
\caption{The way the cubic faces are split into triangular ones in cases
(a) and (b). As far as the external faces are concerned, there is
a $D_{3d}$ symmetry in both cases, but in (a) it is broken by
the internal structure. The symmetry is restorable in the large--lattice
limit by permuting possible internal structures randomly on the lattice.}
\label{figure:cubes}
\end{figure}

In attempting to create easy-to-handle tetrahedral lattices we first
focused on subdividing the elements of a cubic lattice into tetrahedra.
However, there are two essential problems with this:
Firstly, not only cubic edge lengths, but also facial and spatial
diagonals will show up as edges of tetrahedra, making
the correspondence between lattice cell edge--lengths and the
correlation length of the field a poor
one. Secondly, rotational symmetry gets broken, even in a statistical
sense. In ref.~\cite{thesis} we showed the lengthy
proof that there are only two ways of subdividing
a cubic lattice into tetrahedra with matching faces,
and that both subdivisions spoil the rotational invariance of
the large--length limit.
Here we will just present the two possible tetrahedral subdivisions of
cubes (cf.~Fig.~\ref{figure:cubes}):
\begin{itemize}
\item[case (a):]
For this case we have drawn all the tetrahedral
edges except for a spatial diagonal, which can be chosen freely to be any
of the three spatial diagonals {\it not} drawn in that figure.
The drawing (a) in Fig.~\ref{figure:cubes} has a
$D_{3d}= \overline{3}m$ symmetry,
which gets spoiled by the necessary choice of a spatial
diagonal to create the four ``inner'' tetrahedra.
In order for the VV algorithm to work, we need to match
the triangular faces of a cubic cell with the faces of the neighbouring
cells. It therefore makes no difference whether any of
the cubic cells are randomly rotated by multiples of $2\pi/3$ around the
drawn axis. If we allow a lattice to consist of random rotations of 
such cubes, the $D_{3d}$ symmetry can be restored in a statistical sense.
\item[case (b):]
This case is easier to interpret, since all the tetrahedral edges occurring
have actually been drawn. We see that there is again a $D_{3d}$ symmetry
($C_3$ around the drawn axis, plus inversion, plus a mirror symmetry),
but this time this symmetry is manifest even in the cubic unit cell.
\end{itemize}
In either case we end up (at best) with a $D_{3d}$ symmetry, which
singles out a spatial direction. Unsurprisingly, we found that
VV measurements of e.g.~the inertia tensor of string defects on both
lattices have anisotropic Monte Carlo averages.

\begin{figure}[htb]
\centering
\mbox{~}{\hbox{
\epsfxsize=140pt
\epsffile{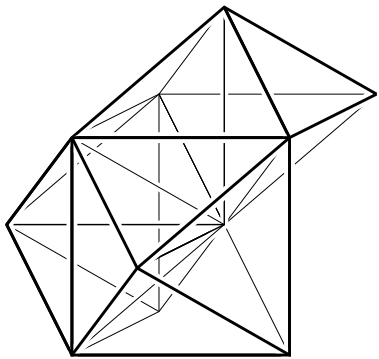}}}\\
\caption{A possible
unit cell of the dual to the
tetrakaidekahedral lattice. There are many ways of
constructing unit cells. This one is elementary and consists of twelve
tetrahedra, which are -- in groups of four --
winding around the connections of the centre of
the cube to the centres of the neighbouring
cubes on top, in front, and to the left of this cube. None of the tetrahedral
faces is part of a cubic face.}
\label{fig:tetrakai}
\end{figure}

\begin{figure}[htb]
\centering
\mbox{~}{\hbox{
\epsfxsize=119pt
\epsffile{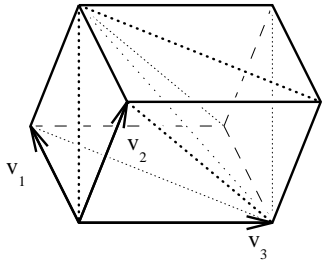}}}\\
\caption{The elementary unit cell of the bcc lattice, with all the
subdivisions to make it the dual of the tetrakaidekahedral lattice.
The base vectors are $\vec{v_1}=\frac{1}{2}(-1,1,1)$,
$\vec{v_2}=\frac{1}{2}(1,-1,1)$, and
$\vec{v_3}=\frac{1}{2}(-1,-1,1)$}
\label{fig:unit-cell}
\end{figure}

However, there exists a tetrahedral lattice
which beats subdivisions of a cubic
lattice in both that it re--establishes rotational symmetry of the
measurables~\cite{MHKS}
and that its edge lengths do not vary by a large ratio~\footnote{This
lattice has also been used by Copeland et.~al.~\cite{Ed}, and Leese and
Prokopec in refs.~\cite{LeePro,LeeProtexture}, for Monte Carlo simulations
of formation of monopoles or textures, and for simulations of
minimally discretised $U(1)$ strings (under the name of tricolor walks)
in ref.~\cite{Bradley}.}~\cite{Scherrer}.
It actually beats the
previously presented ones also in terms of simplicity, because all tetrahedra
are exactly identical. Imagine a body--centred cubic
(bcc) lattice, where all the next nearest and second--nearest
neighbours are thought of as being connected by lattice links. Then this
lattice draws out tetrahedra, all of the same shape, with two edges of
length $a$, the cubic edge length, not touching each other, but being connected
by four edges of length $\sqrt{3}/2\times a$, the nearest
neighbour distance on the bcc lattice.
All edges of the tetrahedra
are nearly equally long and are therefore a reasonable representation of a
given correlation length $\xi$.
This is also reflected in the fact that the
first Brillouin zone of the body-centred cubic lattice,
which builds up the tetrakaidekahedral lattice, is nearly spherical.
We will discuss the dual lattice below.
The tetrahedral lattice is sketched in Fig.~\ref{fig:tetrakai}.
Its point-group symmetries are the same as the ones
of the bcc lattice, i.e.~$O_h$,
and rotational symmetry of measurables is re-established for the Monte Carlo
averages.
The {\it elementary}
unit cell of this lattice, together with it's triangulation,
is drawn in Fig.~\ref{fig:unit-cell}.
We see that this lattice is interpretable as
{\it a simple cubic lattice
subdivided as in type (b), and then tilted triclinically until rotational
symmetry is restored.} The necessary tilt turns
the simple cubic lattice into a body--centred cubic one.
\begin{figure}[htb]
\centering
\mbox{~}{\hbox{
\epsfxsize=190pt
\epsffile{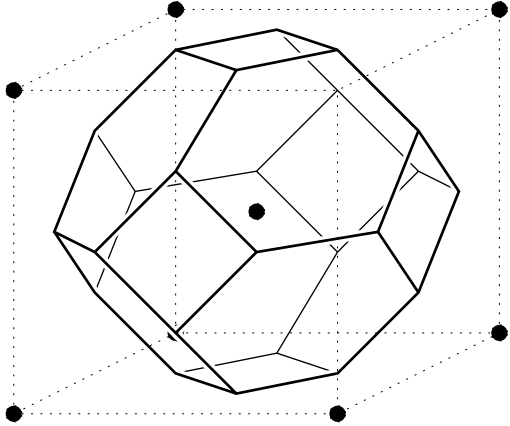}}}\\
\caption{The first Brillouin zone of the bcc lattice, which is also the
unit cell of the tetrakaidekahedral lattice. The field values
are assigned to the centres of those cells (i.e.~to the points of the bcc
lattice), while the strings live on the edges.}
\label{fig:tetrakaiII}
\end{figure}
Fig.~\ref{fig:tetrakai} may be easier to interpret as far as lattice symmetries
are concerned, but Fig.~\ref{fig:unit-cell} is more supportive in
the development of an efficient programming style.

As mentioned already, the unit cells
of the dual lattice are almost spherical. This is desirable because the
random phases assigned to lattice points live in the centres of the elementary
unit cells of the dual lattice (or, more accurately, in the centre of the
first Brillouin zone, which is only one representation of a
minimal unit cell for
the dual lattice). The unit cells of the dual lattice are therefore required
to correspond to a correlation volume, which should obviously be close to
spherical. The dual to our lattice is the tetrakaidekahedral lattice. The unit cell is essentially a cube
with its
vertices cut off half--way between the
centre and the vertex of the cube (cf.~Fig.~\ref{fig:tetrakaiII}).

To have a lattice which is both unique in its
string configurations and rotationally symmetric in its predictions
is an improvement to the VV method which we regard as
highly necessary. Yet, another improvement has been made for the measurements
in \cite{MHKS,MHKSII}: the data structures corresponding to the
string-data allow the lattice to be formally infinite.

\section{The Self--Interacting Walk on an Infinite Lattice}
\label{section:4}

Cosmic strings in the VV algorithm
are a kind of locally self--interacting walk:
There is no need to keep track of vacuum
field values further away from the string than one correlation length.
Effects from other strings are hidden in the randomly assigned vacuum field
values. Every string can be traced from the information in the field
values on the lattice sites. Unless a site has been visited by the {\it same}
string already, we can just assign a {\it random} field value, and
if a site is never visited by the string in question, we need not assign
a field value at all.
 
In effect,
we can ignore the whole lattice, and build up a corresponding data structure
as we need it in the process of tracing the string. This data structure
has to store the coordinates and assigned field values of only those lattice
points which have been needed to tract the string.
We then want
a fast algorithm to check whether a particular lattice point has already
needed an
assignment of a random field value already during previous steps while
tracing the same string, i.e.~a fast way of checking whether given
lattice coordinates are actually present in the data structure or not.
If they are not, a random field value is assigned and the appropriate
values are added to the walk data, otherwise the formerly assigned field
value is re-used and nothing has to be added to the walk data (apart from
whichever measurables one is interested in, of course; they are stored in
a different data structure).

On a finite and storable lattice such a search algorithm is not needed
because we can address each variable in the computer's memory just by the
lattice coordinates, without ever needing to worry whether there are any
other pieces of (the same) string nearby.

Polymer physicists have
studied the SAW, which is meant to describe
configurational statistics  of polymers in a low density solution
\cite{deG,Madras}. To trace out a SAW one has to check at every step whether
the walk has already been at the lattice site which was randomly chosen
to be the next one to be visited by the walk, and an efficient
search algorithm is
an essential ingredient for any method of tracing long
walks~\footnote{``Long" means that the lattice it is expected to stretch
over is too large to be stored as a simple three dimensional array of zeroes
(for lattice sites visited already) and ones (for lattice sites not
visited yet).} \cite{Sokal}.
Naively one would expect the
computational time for this to diverge as the square of the walk
length (or even worse, since the possibility of running into a `dead end'
would have to be avoided at every step, which involves looking ahead
an indefinite number of steps).
However, much more sophisticated methods have become customary in that
field \cite{Sokal}, and the computational
time spent on tracing the walk is usually just
proportional to the walk length, i.e.~the search time for nearby
walk segments is of order one.

In the following sections we will introduce these methods into the
VV type calculations, and argue that hash tables with collision resolution
by double hashing \cite{hashtable} may be the most efficient data
structure for such simulations.

\subsection{A Simple Walk-Array}

Seeing that the rest of the lattice can be ignored, one could store
$L$ data elements in the computer's memory, with $L$ being the total
length of the walk. Every data element stores the coordinates of the
lattice site which was needed to calculate the direction of the appropriate
string segment together with the assigned field value.
When making the next step in the walk, one looks through all the present
array entries. The time it takes to
search whether a given set of coordinates has occurred already in a walk
of length $L$ is of order $L$. The total time it takes to trace such a walk
is therefore proportional to $L^2$. Although the array will usually
be much smaller than having $L$ elements (because of the high coordination
number of our lattice a lattice site can be approached by the walk many
times), this would be bad news if one wants to improve on the measurements
made on finite lattices.
 
\subsection{Classification of Coordinates: a List of Lists}
\label{section:lol}
  
\begin{figure}[htb]
\centering
\mbox{~}{\hbox{
\epsfxsize=300pt
\epsffile{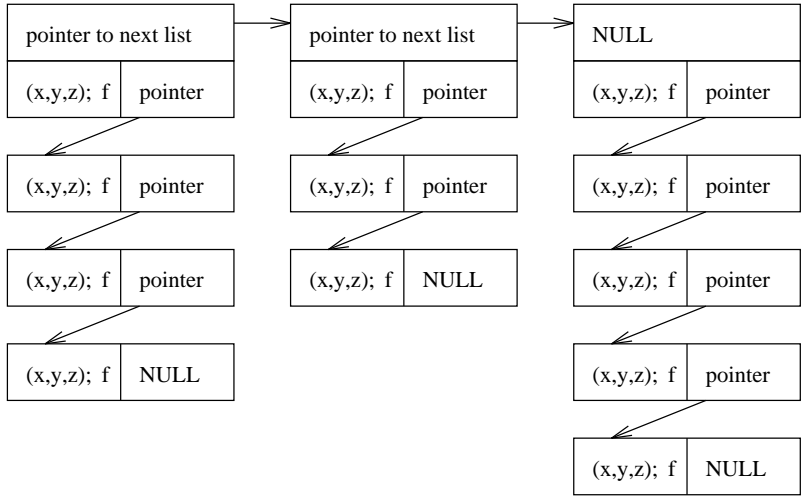}}}\\
\caption{A possible layout of the list of lists after twelve coordinates
have been stored, four of which happened to lie in the first distance
interval from the origin, three lie in the second, and five in the
third. For increased efficiency, the new elements should be stored
{\it on top} of the lists, because they are more likely to be needed
immediately again. More lists are added when the string
moves into the fourth and higher distance intervals.}
\label{fig:listoflists}
\end{figure}

As an improvement to this unstructured array, one can maintain
separate lists for different subsets of the universe of
coordinates. The set of coordinates is then used as a {\it key} from which
we can compute which list we should search in. Of course one would like
the given lists to be approximately equally populated in
order to ensure that the algorithm does not effectively search in one long
list most of the time, while leaving many short lists largely untouched.
Apart from this the criterion by which to classify the universe of
keys is entirely
arbitrary. In \cite{MHKS} we used an algorithm of this kind,
classifying the lattice points by their distance to the origin. The
distance intervals were stacked such that they were expected to be
equally strongly populated by walk elements.
Since one does not know how far a given
walk will distance itself from the origin, it is necessary to link the
data lists dynamically, so that only as many lists as required for every
particular walk are maintained. Every first element of the lists has therefore
in addition to the pointer to the next element a pointer to the first
element of the next list, which is only created when needed. A search
for a given key involves then also a run through all the pointers of the
first elements corresponding to all the distances shorter than the one of
the key to the origin.
A sketch of how the list of lists works is given in Fig.~\ref{fig:listoflists}.
Keeping the length intervals too narrow
slows down the search {\em for} the
right list, whereas keeping them too wide slows down the search
{\it through} the correct list. Therefore the width of these intervals
has to be gauged to achieve a reasonable compromise.
For well--tuned
length intervals, one expects the average search time to diverge as the
square root of the walk length. The time it takes
to trace a walk then diverges proportionally to $L^{3/2}$.
Note that the tuning procedure involves knowing the expected
Hausdorff dimension of the walk, so that
several runs may have to be done with probably non-uniform list lengths
until the Hausdorff dimension is known.

\subsection{Sorted Lists}
\label{section:sortlist}

A better technique has been used already in dynamical simulations
of cosmic strings \cite{EPSS,BB}. There
the lattice volume was divided into small boxes of dimensions comparable to
the average string--string separation, and separate linked lists containing
data for the string segments living within each of these boxes were
maintained. This results initially in an array of lists, many of which
could be empty. However, in the work for ref.~\cite{EPSS}, empty lists were
eliminated from the array, and the remaining lists were sorted by a radix
sort. If we were to use such a method for an infinitely large lattice in
a VV simulation, the time it takes to search {\it for} the right
list is of order $\log (m)$, where $m$ is the number of maintained
(i.e.~occupied) lists. It can easily be shown that search performance
is optimised if $m\sim l$, i.e.~if the average length of an occupied
list is as small as possible. Every lattice site therefore gets its
own box in the ideal adaptation of this algorithm to VV simulations.
The average time of a search is then of order $\log(L)$, but an
{\it unsuccessful} search, i.e.~a search for lattice coordinates which
have not yet been stored, has to be followed by an insertion of the new
walk data somewhere in the array, taking a time of order $L$ to
shift the subsequent array elements. This kind of insertion is not needed
in the work of refs.~\cite{EPSS,BB}, because all searches are performed
at time-slices of the string evolution, i.e.~the whole array of lists is
known from the beginning.
While building up a self-interacting walk, the number of unsuccessful
searches is proportional to $L$, thus the time to trace a string is dominated
by those searches, and will be proportional to $L^2$.

It should be noted that the algorithm of \cite{EPSS,BB} works even
without the elimination of empty lists from the array, because the
lattices are finite. The sort algorithm is essential, however, for
our gradually built-up lattice, such that the time-costly insertions cannot
be eliminated.

\subsection{Hash Tables}
 
The proper adaptation
of the algorithm in section \ref{section:sortlist},
in spite of its other drawbacks for VV type simulations,
eliminated entirely the $L$-dependence of
the search time {\it through} the lists, by storing the data
directly in the array without the use of lists (if we set $m=l$, i.e.~the
box size contains only one lattice point).
How can the $L$-dependence of both the
search {\it for} the right list and the time it takes to
insert a new element be eliminated?

Let us see what the problem in section \ref{section:lol} was.
The origin of the search-time associated with looking {\it for}
the right list was that we didn't know in advance how many lists would
have to be maintained, i.e.~how many classes of keys would be needed
to describe the walk coordinates.

Let us now suppose we have a partition of the universe of keys into
a fixed number of classes of keys.
Then we can write the data into an array of lists, where the correct index
is computed from the key and the appropriate list
can be accessed directly without the use of
pointers. The number of maintainable lists is then
almost arbitrarily large (of order $L$ or larger), and search times through
the lists become of order one.
The prescription of how to compute the index --
giving the position of the list in the array -- from
a given set of coordinates is called the {\it hash function}
$h(x,y,z)$. Given that the set of coordinates one is likely to need
is mapped onto the image of $h$ uniformly, the average search time
needed to find a given set of coordinates which has been stored already
is
\[
\tau=1+\frac{n}{m}\,\,,
\]
where $n$ is the number of
lattice point which have already been stored, and $m$ is the number
of linked lists ($n/m$ is the average length of a list, called
the {\it load factor} $\alpha$ of the hash table).
If the size of the array of lists is comparable to the walk
length, then $\alpha$ is of order one and
the average search time is also of order one.

Such an array of lists
is called a hash table with collision resolution
by chaining \cite{hashtable}.
Collision resolution is the term for the
method that tells us where to keep looking for the key (i.e.~the set of
coordinates of interest) if another key has
been stored in the initially addressed data element. Chaining is the
name for this method, if the place where we keep looking is a dynamically
linked list.

\subsection{Collision Resolution by Open Addressing}

There is an even more efficient method for our
purposes, which avoids the use of pointers altogether. It can only
be used if no data entry has to be deleted until the whole hash table
can be erased, but this is the case when one traces a static walk.
In collision resolution by open addressing, every data element is stored in
the original array, the hash table $T_{im}$, where $i$
is the index that counts the
lattice coordinates and other variables to be stored for every point
on the walk, and $m$ is of the order of, but larger than, the maximum
length of any walk we would like to trace.
First we compute the array index from the set of
coordinates, using a hash function as the arbiter of the class partition
of lattice points. If $U$ is the universe of keys -- or the set or lattice
sites -- the hash function maps
\begin{equation}
h:\,U\rightarrow\{0,\ldots,m-1\}\,\,\,\,\,,\,\,\,\,\,
\label{eq:hash}
\end{equation}
and appropriate variables for the key $k=(x,y,z)$, plus they key itself,
will be stored in the
slot $T_{i,h(k)}$ of the hash table $T$.
Any $T_{i,}$ therefore contains $x,y,z$ and some set of variables
$f(x,y,z)$.
Obviously, collisions will occur for certain keys,
i.e.~$h(k_1)=h(k_2), \mbox{~for~} k_1\neq k_2$.
These can only be identified because the key itself has also
been stored in the slots of $T$.
To make collisions as unlikely as possible, one should make sure that
all $h(k)$ are expected to be uniformly distributed
over $\{0,\ldots,m-1\}$, and keep a low load factor $\alpha$.
The time it takes to find a given set of coordinates in the hash table
then depends on $\alpha$ and the method used for collision resolution.
If we know that we can keep $\alpha<1$ and that we do not have to delete
single elements from the hash table,
collision resolution is best achieved by so-called {\it open
addressing}, which gives us a rule by which to find an alternative
slot in the hash table,
if the slot we are searching in is already occupied by another key and
its field data.
Then we select a new $m$ which might still be empty,
instead of appending a list of elements to the slot $T_{,m}$

\subsection{Double Hashing}

There are many ways of selecting alternative slots (cf.~ref.~\cite{hashtable}).
The one we used in \cite{MHKSII} is called {\em double hashing}:
One uses two hash functions
$h_1 , h_2:\,U\rightarrow\{0,\ldots,m-1\}$. The search algorithm looks
first at the slot $T_{i,h_1(k)}$. If this slot is
occupied by a different key from the one
we are looking for, it looks
at the slot with index $(\left[h_1(k)+h_2(k)\right]\bmod m)$,
and so forth through all the
$(\left[h_1(k)+nh_2(k)\right]\bmod m)$
for $n=0,\ldots,m-1$.
If one encounters an empty slot
before finding the key, then the key simply has not been used yet.
A search in which the key is not
found (since it has not been stored yet) is called an unsuccessful search.
Insertion of new data is therefore preceded by an unsuccessful search, and
the data is inserted at the first empty slot found.
For collision resolution by double hashing the average time of an
unsuccessful search is at most $1/(1-\alpha)$, while the average time
of a successful search is at most
$\frac{1}{\alpha}\left(1+\ln\left(1/(1-\alpha)\right)\right)$,
but in any case less
than the time for the
average unsuccessful search (since it corresponds to an unsuccessful
search at the time the given key was inserted, which was always at a time
of a lower load factor), given that two
restrictions on the choice of hash functions apply \cite{hashtable}:
\begin{itemize}
\item
\raggedright
The set of $h_1(k),\,k\in U$ is uniformly distributed in
$\{0,\ldots,m-1\}$.\\
\raggedleft
condition (1)
\item
\raggedright
$h_2(k)$ must be relatively prime to $m$ for all $k$, so that
the whole hash table is searched if no empty slot can be found, i.e.\\
$\{\left[h_1(k)+nh_2(k)\right]\bmod m\, |\,n\in \{0,\ldots,m-1\}\}
\equiv \{0,\ldots,m-1\}$.\\
\raggedleft
condition (2)
\end{itemize}
Reasonable compromises between the desire for good memory
utilisation and computational speed can
be reached at load factors of about $1/2$ or $1/3$, although higher load
factors are possible. One can gauge the desired compromise between
efficient memory-utilisation and speed better than one can while using
chaining for collision resolution, since the (essentially unknown) number
or data collisions do not claim additional memory. Moreover, the absence of
pointers reserves more memory for data in any case.
Double hashing seems
to be the most efficient method of open addressing. Others are linear and
quadratic probing, which search the sequences
$(h(k)+n){\rm mod}\,m, \,n=0,\ldots,m-1$
and $(h(k)+n+cn^2){\rm mod}\,m,\,n=0, \ldots,m-1$
respectively. For linear probing one gets clusters of
occupied slots, which increase the probability of a further data collision, if
one collision has occurred already in the search, and for quadratic probing it
is generally quite difficult to ensure that all slots are searched through.

\subsection{Example: Double Hashing on a bcc Lattice}

It is now straightforward to apply hash tables with double hashing in
VV-type algorithms.
Remember that the assignment of field values happens
randomly when the string comes into
the vicinity of a lattice point for the first time, but that
these values have to
be recalled if the string approaches this point again at any later time.
Using a hash table, we can simulate
an arbitrarily large lattice, without considering more points than necessary.
One can easily trace strings with lengths of up to $10^7$ lattice units
on a small workstation without prohibitive memory
requirements. This is a necessity for the accurate determination of the
statistical measurables of e.g.~the percolation model of string
statistics in ref.~\cite{MHKSII}.
  
It is particularly easy to create a uniformly
distributed $h_1(k)$ on a simple cubic lattice $\Z_3$. For example,
we could set $m$ to be of the form $m=s^3$,
so that the set of all slots corresponds to the elements of
$\Z_s^3$, and cubes of edge-length $s$ can be stacked and mapped
identically onto $\{0,1,\ldots,m-1\}$
\begin{equation}
h_1(x,y,z)=\left[x \bmod s + s(y \bmod s) + s^2(z \bmod s)\right]\,\,\,\,\,.\,\,\,\,\,
\label{eq:hash1}
\end{equation}
Condition (2) is
fulfilled if $h_2(k)$ is relatively prime to $s$. This can be achieved
by making $s$ a power of two, and ensuring that $h_2(k)$ is odd, for example
\begin{equation}
h_2(x,y,z)=\left\{2\left[y + s z  + s^2 x\right]+1\right\} \bmod m\,\,\,\,\,.\,\,\,\,\,
\label{eq:hash2}
\end{equation}
Note that $h_2(k)$ does not map every
cube of size $s^3$ onto $\{0,\ldots,m-1\}$
in identical ways, so that coordinates that are identical modulo $s$
do have the same starting
slot $T_{h_1}$, but a different search-sequence thereafter.
Since $h_2$ has to be relatively prime to $m$, the uniformity of $h_2$ is
somewhat restricted, but on average we still map onto all odd numbers $<m$
equally often.

Obviously, we cannot use Eq.~\ref{eq:hash1} on the body-centred cubic
lattice,
since all the points have either only even or only odd coordinates, and
$h_1$ would not be very uniform.
We therefore create a unique map from the body-centred cubic
lattice to a simple cubic
one, and use as hash functions
\begin{equation}
g(x,y,z)=h\left(\left[\frac{x+1}{2}\right],\left[\frac{y+1}{2}\right],
z\right)\, ,
\label{eq:map}
\end{equation}
with $h$ being defined as $(h_1+nh_2)\bmod m$, and $n$ the number of data
collisions that had to be resolved to search for the desired slot in $T$.
The Gauss bracket $\left[a\right]$ denotes the nearest
integer not larger than $a$.

The hash functions Eqs.~\ref{eq:hash1},\ref{eq:hash2},
and \ref{eq:map} are the ones used in ref.~\cite{MHKSII}. There, this algorithm
allowed us to investigate a known percolation transition in string
defect statistics \cite{MHKS} much closer than it is possible
on a finite lattice. The transition
was postulated in \cite{V91}, but before the infinite lattice view
it has been impossible to measure critical exponents of the transition
appropriately.
We found that the critical exponents of the
percolation transition are in the same universality class as -- and
therefore identical with -- standard bond or site percolation exponents.
We were also able to measure the fractal dimension
of string defects to higher accuracy, and to observe subtle differences
between strings
of only slightly different symmetry groups, while transforming $U(1)$
strings smoothly into $\R P^2$ strings.

Not specifying a lattice boundary also eliminates dramatic finite size
errors, which are both very clearly observable \cite{Allega} and well
understood theoretically \cite{Aus+}.

\section*{Conclusion}

We have discussed necessary consistency requirements and improvements
to the Vachaspati Vilenkin algorithm for Monte Carlo measurements
of the statistics of cosmic strings and line disclinations.

We have shown a simple criterion which ensures that VV type simulations
preserve the topological definition of the string flux through a given
surface in the lattice description. We have suggested the use of a tetrahedral
lattice to solve problems of uniqueness present
on cubic lattices, although there are symmetry groups which
escape a unique determination of the string shapes even on those
lattices. We have argued that the lattice which is dual to the
tetrakaidekahedral lattice combines further two desirable properties:
it preserves the rotational symmetry of Monte Carlo averages, and it has
only slightly varying edge lengths, making their correspondence to a
well-defined physical correlation length more sensible.
The first Brillouin zone of
the dual lattice, which corresponds to a correlation volume after the
string-forming phase transition, is close to spherical.

We have introduced the possibility to view string defects as locally
self-interacting walks, allowing us to use a formally infinite lattice,
which is built up as one needs it to construct the defects, and avoiding
the specification of any boundary conditions.

This requires an efficient search algorithm to look up phase
values on lattice sites lying on lattice plaquette which have been
pierced by the string already. We find that a hash table algorithm
with collision resolution by open addressing allows us to search for
such points in a computational time of order one, permitting to
trace a string in the time it takes to only write down the walk
coordinates. Typically, a computer with 128MB of RAM will allow to
trace strings of up to $10^7$ lattice units, which improves on measurements
on finite lattices by two to three orders of magnitude.

\section*{Acknowledgments}

This work is supported by PPARC Fellowship GR/K94836.

\end{document}